\documentclass[11pt,a4paper]{article}

\usepackage[margin=16 pt, labelfont=bf]{caption}
\usepackage{subcaption}
\usepackage{siunitx}
\usepackage{mathtools}
\usepackage{bbm}

\usepackage{graphicx}%
\usepackage{multirow}%
\usepackage{amsmath,amssymb,amsfonts}%
\usepackage{amsthm}%
\usepackage{mathrsfs}%
\usepackage[title]{appendix}%
\usepackage{xcolor}%
\usepackage{booktabs}%
\usepackage{longtable}%
\usepackage{listings}%
\usepackage{geometry}

\usepackage{hyperref}
\usepackage{cleveref}

\hypersetup{
    colorlinks=true,       
    linkcolor=black,       
    citecolor=black,       
    urlcolor=blue,         
    allcolors=blue         
}

\DeclareMathOperator*{\argmax}{arg\,max}

\title{{\bf Bayesian Inversion via Probabilistic Cellular Automata: an application to image denoising}}
         
\author{ {\bf Danilo Costarelli}, {\bf Michele Piconi}, {\bf Alessio Troiani} \\  
Department of Mathematics and Computer Science \\
            University of Perugia\\
        1, Via Vanvitelli, 06123 Perugia, Italy    \\  
{\small {\tt danilo.costarelli@unipg.it}} - {\small {\tt michele.piconi@unipg.it}} \\
{\small {\tt alessio.troiani@unipg.it}}
}

\begin{document}

\maketitle


\begin{abstract}
We propose using Probabilistic Cellular Automata (PCA) to address inverse problems with the Bayesian approach. In particular, we use PCA to sample from an approximation of the posterior distribution. The peculiar feature of PCA is their intrinsic parallel nature, which allows for a straightforward parallel implementation that allows the exploitation of parallel computing architecture in a natural and efficient manner.
We compare the performance of the PCA method with the standard Gibbs sampler on an image denoising task in terms of Peak Signal-to-Noise Ratio (PSNR) and Structural Similarity (SSIM). The numerical results and the large speedups obtained with this approach suggest that PCA-based algorithms are a promising alternative for Bayesian inference in high-dimensional inverse problems.

\vskip0.2cm
\noindent {\em AMS subject classification:} 60J22, 62F15, 62M20, 62M40
\vskip0.2cm
\noindent {\em Key words:} Bayesian inversion; Markov Chain Monte Carlo, Probabilistic Cellular Automata; Gibbs sampler; image denoising
\end{abstract}


\section{Introduction}\label{sec:introduction}

Loosely speaking, an inverse problem consists of determining the \emph{root cause}
of an observed phenomenon. Several examples fall into this category, spanning a wide range of fields.

Problems of this type arise very frequently in remote sensing, for instance, in the context of large-scale monitoring of Essential Climate Variables (ECVs) such as Soil Moisture (SM) \cite{balenzano2010dense, clarizia2019analysis}, Freeze-Thaw state (FT) \cite{mcdonald200553, rautiainen2021freeze}, and Above Ground Biomass (ABG) \cite{bouvet2018above, chen2018estimation}.
In this context, it is common to have two-dimensional arrays of some measured backscattered signal that one wants to convert into levels of some measured physical quantity, such as soil permittivity, and, in turn, to quantitative values of the ECV of interest. Note that the measured signal will, in general, be affected by noise. Problems like this are typically ill-posed, and several strategies have been developed to tackle them (see \cite{angeloni2024microwave, mereu2025interpolation, veneri2025review} for comprehensive reviews).

Among such strategies, we consider the Bayesian approach to retrieval (as in \cite{kerr2012smos, paloscia2008comparison}), which has the peculiar feature of treating all quantities at stake as random variables. The key object of the Bayesian approach is the so-called \emph{posterior distribution}, which is a probability distribution of the possible root causes given the observed data. The retrieval is then achieved by sampling from the posterior distribution, typically by collecting samples from a Markov chain with the posterior distribution as its stationary distribution. This approach is known as the Monte Carlo Markov Chain (MCMC). For a comprehensive treatment of Markov chains and MCMC methods, the reader can refer to \cite{bremaud2020markov}.

In this paper we propose an MCMC approach where the considered Markov Chain is a Probabilistic Cellular Automaton (PCA), that is a Markov Chain that at each step, updates all components of the state vector independently (conditionally, given the past) one from another (\cite{cirillo2022metastability, cirillo2008metastability, fernandez2018overview, lebowitz1990statistical}) with a probability that is affected only by the \emph{local} effect of the update of each component, that is an update probability that, for a given site, only depends by the state at the neighboring sites.
Due to its parallel nature, a PCA can be simulated effectively on a parallel computing architecture such as a GPU (\cite{lancia2013equilibrium}) where, in principle, each computing core may take care of the update of a single component of the state vector. All these updates may be performed simultaneously.
In general, it is not immediate to define a chain of this type having the stationary distribution equal to the desired posterior distribution. However, we will see how it is possible to define a PCA whose stationary distribution is close to the desired one (see, e.g., \cite{dai2012sampling}).

Our aim is to show that the PCA approach can be a viable alternative to the use of other MCMC approaches, such as the Gibbs Sampler and the Metropolis-Hastings algorithm, where to update a subset of the components of the state space, the \emph{global} effect of the update must be evaluated. Since evaluating this global effect can be computationally expensive, it is common in the literature to consider only single-component sequential updates. A similar strategy, though in a different context, has been used, e.g., in \cite{zaheer2016exponential}.

To support the viability of our approach, we consider simple cases of noisy image restoration. To this end, we investigate one of the scenarios discussed in the seminal paper \cite{geman1984stochastic}, which greatly contributed to the popularity of the Bayesian approach to inversion.

This paper aims to evaluate the effectiveness of PCA in addressing inverse problems within the Bayesian framework. In this respect, the results presented here are rather promising.
In particular, we show that the PCA is able to produce slightly better results with respect to the Gibbs sampler in terms of Peak Signal-to-Noise Ratio (PSNR) and Structural Similarity Index (SSIM) while reducing the execution times drastically.

In \Cref{sec:background}, we provide some background on the Bayesian approach to inversion and describe how Markov Chain Monte Carlo (MCMC) techniques can be used to draw samples from the posterior distribution. In particular, in this section, we recall the definition of the Gibbs Sampler and establish the relationship between its stationary measure and the posterior distribution.

In \Cref{sec:probabilistic_cellular_automata} we give the definition of Probabilistic Cellular Automata, characterize their stationary measure, and show how this distribution can be directed towards the desired posterior distribution.
In \Cref{sec:image_model}, we describe the prior distribution of the test images, whereas in \Cref{sec:retrieval_algorithms}, we present the details of the retrieval algorithms we consider.
In \Cref{sec:experimental_results}, we provide a comparison of the results obtained with the PCA approach with those of the Gibbs sampler in terms of Peak Signal to Noise Ratio (PSNR) and Structural Similarity Index (SSIM) and execution times.
Finally, in \Cref{sec:discussion} we highlight future lines of research.


\section{Background}\label{sec:background}

\textbf{Bayesian Inversion.}
An inverse problem concerns the estimation of an object of interest $x$ from an observed or measured quantity $g$. The relationship between $x$ and $g$ is given in terms of a mathematical model
$$
g = \psi(x, \varepsilon)
$$
where $\varepsilon$ is a term that accounts for noise affecting the observation, as well as other unobserved or poorly known quantities.
In this context, solving an inverse problem amounts to finding an ``inverse'' of the function $\psi$.

Both $x$ and $g$ may be quite general objects. For example, in the context of remote sensing, $x$ could refer to soil permittivity, which is related to essential climate variables such as Soil Moisture and Freeze-Thaw state, while $g$ might be the measured backscatter of a signal.
In the context of image processing, $x$ may be the ``true'' image of interest, whereas $g$ represents the degraded observed version of that image affected by factors such as noise (the $\varepsilon$ term) or blur (a suitable form of $\psi$).

In many applications, the problem of inverting $\psi$ is not well posed, and several approaches have been developed to tackle it (see \cite{aster2018parameter} for an overview in the context of parameter estimation and \cite{bertero2021introduction} for an introduction in the field of imaging).
Here we consider the so-called Bayesian approach (for a comprehensive introduction, see the monograph \cite{kaipio2004statistical}).
In the Bayesian framework, all quantities at stake are treated as (in general, multivariate) random variables linked by the relation
$$
G = \psi(X, N),
$$
where \( G \) denotes the observed data, \( X \) represents the values to retrieve, and \( N \) stands for the noise.
In this framework, the inverse problem consists of determining the so-called posterior distribution
$$
\pi_{\mathrm{post}}(X = x | G = g).
$$
In the following, we always denote random variables with capital letters, whereas lowercase letters will denote the values taken by the (usually corresponding) random variables.
When it does not give rise to ambiguity, we do not write the name of the random variable explicitly and write expressions as the previous one as $p_{\mathrm{post}}(x|g)$.
By Bayes' theorem
$$
\pi_{\mathrm{post}} (x | g) = \frac{\pi(g | x) \pi_{\mathrm{prior}}(x)}{\pi(g)}
$$
where $\pi$ is the joint distribution of $X$ and $G$ and $\pi_{\mathrm{prior}}$, called the prior distribution of $X$, models the knowledge we have on the quantity we want to estimate \emph{before} the outcome of the experiment (measured physical quantity, noisy image, ...) has been observed.

Here, we only consider models of the form
\begin{equation}\label{eq:general_forward_model}
G = \phi(X) \odot N
\end{equation}
where $\odot$ is an invertible function such as addition or multiplication. Furthermore, we always assume that $N$ and $X$ are independent random variables. Denoting by $\Phi$ the inverse of $\odot$, we have 
\begin{equation}\label{eq:likelihood_general_expression}
\pi(g | x) = \pi_{\mathrm{noise}} \{N = \Phi(g, \phi (x)) \}
\end{equation}
which is called the \emph{likelihood} of the observed data given the \emph{ground truth} \( x \).
To clarify the interpretation of \( \Phi \) in the previous formula, if \( \odot \) denotes addition, then \( N = \Phi(g, \phi(x)) \) corresponds to \( N = g - \phi(x) \).

Observing that, conditional on $G = g$, the marginal density $\pi(g)$ is a constant, it is easily seen that
$$
\pi_{\mathrm{post}}(x | g) \propto \pi_{\mathrm{prior}}(x) \pi_{noise}(\Phi(g, \phi(x)))
$$
where \( \propto \) means ``proportional to''.
The posterior distribution is, therefore, determined by the prior probability assumed for $X$ and the model chosen to describe the noise.

Note that we provide the expression of \( \pi_{\mathrm{post}}(x | g) \) only up to a multiplicative factor that would turn the previous expression into a proper probability density. In the previous case, this multiplicative constant is the marginal density \( \pi(g) \). In general, this normalizing factor is hard to evaluate explicitly. However, using the so-called Markov Chain Monte Carlo (MCMC) approach (see below), it is possible to sample from probability distributions only known up to a multiplicative factor.
In what follows, we often write un-normalized probability densities (using the symbol \( \propto \)) when the knowledge of the normalizing constant is not needed in the retrieval process.

In this paper, we consider prior distributions of the form
\begin{equation}\label{eq:gibbs-prior}
\pi_{\mathrm{prior}}(x) = \frac{ e^{- \beta H(x)}}{Z}.
\end{equation}
A distribution of this type is called a \emph{Gibbs} distribution and originates in statistical mechanics. The parameter \( \beta > 0 \) is called \emph{inverse temperature}.
The \emph{energy} function $H$ is called Hamiltonian, and $Z$ is 
the normalizing constant (referred to as \emph{partition function} in the context of statistical mechanics). Note that asking for a prior of the form \eqref{eq:gibbs-prior} amounts to requiring that every state \( x \) has strictly positive prior probability. Indeed, if the prior probability of \( x \) is proportional to a positive weight \( w(x) \), the corresponding prior probability is, up to a normalizing constant, \( e^{ \log w(x) } \).

As for the likelihood, we consider (additive) Gaussian noise with mean $\mu$ and variance $\sigma^{2}$.
Then, the likelihood is of type
\begin{equation}
\pi_{\mathrm{noise}}(g | x) \propto e^{-\frac{1}{2\sigma^{2}}\|\mu - \Phi(g, \phi(x)\|_{2}^{2}}
\end{equation}
and the posterior distribution, therefore, of type
\begin{equation}\label{eq:gibbs_posterior}
    \pi_{\mathrm{post}}(x | g) 
    \propto e^{- \beta \left[ H(x) + \frac{1}{\beta}\frac{1}{2\sigma^{2}}\|\mu - \Phi(g, \phi(x)\|_{2}^{2} \right]}.
\end{equation}
That is, it is again a Gibbs distribution with Hamiltonian
\begin{equation}\label{eq:posterior_hamiltonian}
    H_{g}(x) = H(x) + \frac{1}{\beta}\frac{1}{2\sigma^{2}}\|\mu - \Phi(g, \phi(x)\|_{2}^{2}.
\end{equation}
In this paper, we refer to the Hamiltonian defining the posterior Gibbs distribution as the \emph{posterior Hamiltonian}.

\textbf{Bayesian inference.}
In a deterministic framework, the inversion process aims to produce a single ``recovered'' value. On the other hand, in the Bayesian framework, the outcome of the inversion process is the posterior distribution, which is a probability distribution over the entire set of values that can be taken by the variable of interest $X$. However, to obtain useful information from the posterior distribution, we have to specify how this information must be ``extracted'' from this distribution. Typical choices are the so-called \emph{conditional mean estimate} (CM) and \emph{maximum a posteriori estimate} (MAP), both providing a point estimate for the quantity to be recovered.
The former is defined as
\begin{equation}
\hat{x}_{\mathrm{MAP}}:=\argmax_{x}\left\{\pi_{\mathrm{post}}(x | g)\right\},
\end{equation}
that is, it is the value that maximizes the posterior density,
whereas the latter is
\begin{equation}
\hat{x}_{\mathrm{CM}}:=\operatorname{\mathbb{E}}(X|g)=\int x \cdot \pi_{\text {post }}(x \mid g) \mathrm{d} x,
\end{equation} that is, the average value with respect to the posterior density of the quantity one wants to recover.

Although, in principle, computing the maximum a posteriori and conditional mean estimates is straightforward, in many real situations, solving the optimization problem or computing the integral may be a challenging task. This is due to the fact that the problem lives in a high-dimensional space. A common scenario is \( x \in S^{n} \) where \( S \) is a finite alphabet. As an example, in the case where \( x \) represents a grayscale image, \( S \) is the number of allowed gray levels for each pixel, and \( n \) is the number of pixels. Even for pure black-and-white images of very limited size (e.g., \( 32 \times 32 \)), the state space is still huge.

Both the optimization and integration problems can be solved using a Monte Carlo (MC) approach. 
For the computation of the conditional mean, it is possible to exploit the law of large numbers. Indeed, consider a random variable \( X \in \mathcal{X} \) and let \( x_{1}, x_{2}, \dots, x_{K} \) be a collection of points drawn from the probability distribution \( f_{X} \) of $X$. Then, as $K$ gets large,
\begin{equation}
\operatorname{\mathbb{E}}h(X) = \int_{\mathcal{X}} h(x) f_{X} \operatorname{d}x \approx \frac{1}{K} \sum_{i=1}^{K} h\left(x_{i}\right).
\end{equation}
On the other hand, to solve the optimization problem, it would be possible to draw samples from the neighborhood of the maximizers of \( f_{X} \).

Consequently, a key point to effectively perform Bayesian inversion is the ability to collect samples from the posterior distribution. One way to achieve this task is to consider the 
MCMC approach: 
define a Markov Chain $X^{(t)}$ living on \( \mathcal{X} \) admitting a prescribed distribution \( \pi \) as invariant measure. Then, under mild conditions on \( X^{(t)} \), (irreducibility and aperiodicity), the limiting distribution of the chain, also called the stationary measure, is \( \pi \), independent of the initial state of the chain (see \cite{haggstrom2002finite} for a proof).

When the state space of the chain is \( \mathcal{X} = S^{n} \), one possibility to have a prescribed \( \pi \) as invariant measure is to consider the Gibbs-Sampler algorithm. 
According to this sampling scheme, the i-th component of \( X \) is updated according to the conditional distribution \( \pi ( X_{i} = \cdot | \{ X_{j} = x_{j}, j \neq i \} ) \). More properly:
\begin{multline}\label{eq:single_component_gibbs_sampler_update_probability}
    \operatorname{\mathbb{P}}\left( X^{(t+1)} = (x^{(t)}_{1}, \dots x^{(t)}_{i-1}, x_{i}, x^{(t)}_{i+1}, \dots x^{(t)}_n ) \middle| X^{(t)} = (x^{(t)}_{1}, \dots x^{(t)}_{n}) \right) 
\\ = \pi\left( X_{i} = x_{i} \middle| \{ X_{j} = x^{(t)}_{j}, j \neq i \} \right).
\end{multline}
The component to be updated may be chosen at random, or the components may be updated systematically, e.g., sequentially (this is called the systematic Gibbs sampler). In analogy with the statistical mechanics parlance, we call this sampling scheme the (systematic) \emph{single-spin-flip} Gibbs sampler.

Consider the case where \( \pi \) is a Gibbs distribution with Hamiltonian \( H \) over the space \( S^{n} \). 
If \( H \) has the form \( H(x) = \sum_{i = 1}^{n}  f_{i}(x, x_{i})\), then the conditional probabilities \( \pi(X_{i} = s | \{X_{j} = x_{j}, j \neq  i\}) \) are of type
\begin{equation}
 \pi(X_{i} = s | \{X_{j} = x_{j}, j \neq  i\}) = \frac{e^{ \beta f_{i}(x, s) }}{\sum_{s \in S} e^{ \beta f_{i}(x, s) }}.
\end{equation}

Their computation is, therefore, straightforward (at least in the case where \( S \) is ``small'').
Further, it is common that \( f_{i} \) only depends on the values of \( x \) in a \emph{neighborhood} of component \( i \) (e.g., the pixels that are close to pixel \( i \) in the case of image) and, hence, its computation is fast even if the number of components is large.

Note that, usually, the interesting time scale of the chain is not the scale where a single update takes place, but, rather, the scale defined by \emph{sweeps}, that is, sequences of $n$ updates. Therefore, in a sweep, the systematic Gibbs sampler samples a new value for each component of \( x \).

\textbf{The Geman-and-Geman paper.}
In their seminal paper \cite{geman1984stochastic}, Geman and Geman show that in the case of Gibbs prior distribution and noise independent of the ``true'' signal, in the limit \( \beta \to \infty \), the stationary measure of the Gibbs Sampler is the uniform distribution over the maxima of the posterior distribution, provided the cooling schedule is ``slow enough''. 
In particular, they show that the result holds if the inverse temperature \( \beta \) goes to infinity as \( \log k \), where k is the number of steps of the chain. Unfortunately, this cooling scheme is too slow for practical applications. However, their results demonstrate that the MCMC approach can be a viable option for tackling optimization problems. Furthermore, their successful application of the Gibbs Sampler to image restoration demonstrated that even practically feasible cooling schemes can be beneficial in several applications.


\section{Probabilistic Cellular Automata}\label{sec:probabilistic_cellular_automata}

In principle, for the Gibbs sampler to work, it is not necessary to update a single component of \( X_{i} \) at a time. Consider \( X \in S^{n} \) (that is, \( X = (X_{1}, X_{2}, \dots X_{n}) \)) and let \( \pi \) be a probability distribution on \( S^{n} \). 
Let \( I \subset \{ 1, 2, \dots, n \} \) and call \( J = \{ 1, \dots n \} \setminus I \).
Then, a Gibbs sampler is a Markov chain whose transition probabilities are given by
\begin{multline}\label{eq:multi_component_gibbs_sampler_update_probability}
    \operatorname{\mathbb{P}}\left( X^{(t+1)} = (x_{i}, \, i \in I; x^{(t)}_j, j \, \in  J) \middle| X^{(t)} = (x^{(t)}_{1}, \dots x^{(t)}_{n}) \right) 
    \\ = \pi\left( \{ X_{i} = x_{i}, \, i \in I \} \middle| \{ X_{j} = x^{(t)}_{j}, \,  j \in J \} \right).
\end{multline}
Also in this case, there is some freedom in choosing the set \( I \). One possibility is to fix the size of \( I \) to \( k \) and, at each step, sample a new subset of indices to be updated of size \( k \).

However, in general, the computation of the conditional probabilities in \eqref{eq:multi_component_gibbs_sampler_update_probability} may be highly demanding from a computational point of view when \( I \) is not a singleton.

A drawback of the single spin flip Gibbs sampler is the fact that the updates of two different components of \( X \) are, in general, not independent. This dependence does not allow for the exploitation of the computational capabilities of parallel processors (such as GPUs) to their fullest.
It would be tempting to update all components of \( X \) \emph{in parallel} and perform a sweep in a single pass, that is, set\footnote{Note that here (and in what follows), we use the capital letter \( T \) to denote the generic time step of the chain in place of the lower case letter \( t \). This is to highlight that with this update rule, the timescale is that of a full \emph{sweep:} all components have a chance to be updated.}
\begin{multline}
\operatorname{\mathbb{P}}\left( X^{(T+1)} = (x_{1}, \dots x_{n}) \middle| X^{(T)} = (x^{(T)}_{1}, \dots x^{(T)}_{n}) \right) 
    \\ = \prod_{i = 1}^{n} \pi(X^{(T+1)}_{i} = x_{i} | X^{(T)} = (x^{(T)}_{1}, \dots x^{(T)}_{n})).
\end{multline}
Unfortunately, this update rule does not have the desired measure \( \pi \) as the stationary distribution (for a characterization of the stationary measures of PCA see \cite{dai2002stationary} and references therein). Further, its long-term behavior may be highly dependent on the initial condition.

A Markov Chain on \( S^{n} \) with transition probability matrix defined as
\begin{equation}\label{eq:general_pca_definition}
\begin{aligned}
\operatorname{P}(x, w) 
    & \coloneq \operatorname{\mathbb{P}}\left( X^{(T+1)} = (w_{1}, \dots w_{n}) \middle| X^{(T)} 
        = (x_{1}, \dots x_{n}) \right) \\
    & = \prod_{i = 1}^{n} \operatorname{\mathbb{P}}(X^{(T+1)}_{i} = w_{i} | X^{(T)} = x)
\end{aligned}
\end{equation}
is called a Probabilistic Cellular Automaton (PCA). 
A Markov Chain of this kind can be effectively simulated on a parallel computing architecture. This is because, given the current configuration, the value of each component can be sampled independently from the others. A dedicated computing unit can determine each new value.

The PCA that we describe in the following sections have a stationary distribution that is 
\emph{close} to the actual posterior distribution. This feature is achieved by introducing an inertial term in the Hamiltonian, which prevents too large changes at each step of the chain. Although this sampling scheme yields a stationary distribution that is only an approximation to the desired posterior distribution, the increased computing efficiency makes this approach worth further investigation.

Consider a Hamiltonian \( H(x) = \sum_{i = 1}^{n} f_{i}(x, x_{i}) \) and let \( X \) be a Markov Chain on \( S^{n} \) with transition probability matrix
\begin{equation}\label{eq:double_hamiltonian_pca_general_form}
\operatorname{P}(x, w) = \frac{e^{ -\beta H(x, w) }}{\sum_{w} e^{ -\beta H(x, w) }}
\end{equation}
where $H(x, w) = \sum_{i} f(x_{i}, w_{i})$ is a \emph{lifted} two-arguments version of the original Hamiltonian \( H \).
Then it is immediate to show that \eqref{eq:double_hamiltonian_pca_general_form} defines a PCA in the sense of \eqref{eq:general_pca_definition}. 
Indeed,
\begin{equation}\label{eq:double_Hamiltonian_PCA}
\operatorname{P}(x, w) 
    \propto e^{ -\beta H(x, w) } 
    = e^{ - \beta \sum_{i = 1}^{n} f_{i}(x, w_{i}) }
    = \prod_{i = 1}^{n} e^{ -\beta f_{i}(x, w_{i}) }.
\end{equation}
We have
\begin{equation}
\operatorname{P}(X^{(T+1)}_{i} = w_{i} | X^{(T)} = x) = \frac{\operatorname{\mathbb{P}}( X^{(T+1)} = w_{i}, X^{(T)} = w)}{\operatorname{\mathbb{P}}(X^{(T) }= x)}
\end{equation}
Note that
\begin{equation}
\begin{aligned}
\operatorname{\mathbb{P}}(X^{(T)} = x) 
    & = \sum_{w} \operatorname{\mathbb{P}} (X^{(T+1)} = w | X^{(T)} = x) \\ 
    & \propto \sum_{w_{1} \in S} \dots \sum_{w_{n} \in S} \prod_{j = 1}^{n} e^{ -\beta f_{j}(x, w_{j})} \\ 
    & = \prod_{j=1}^{n} \sum_{w_{j} \in S} e^{ -\beta f_{j}(x, w_{j})}.
\end{aligned}
\end{equation}
Similarly,
\begin{equation}
\operatorname{\mathbb{P}}( X^{(T+1) = w_{i}}, X^{(T) = w}) 
    \propto e^{ -\beta f_{i}(x, w_{i}) } \prod_{j  \neq i} \sum_{w_{j} \in S} e^{ -\beta f_{j}(x, w_{j})}.
\end{equation}
Then the claim follows with
\begin{equation}
\operatorname{\mathbb{P}}(X^{(T+1)}_{i} = w_{i} | X^{(T)} = x)
    = \frac{e^{ -\beta f_{i}(x, w_{i}) }}{\sum_{s \in S} e^{ -\beta f_{i}(x, s) }}.
\end{equation}

Further note that, if \( H(x, w) = H(w, x) \), then it is easy to show that the stationary  measure of this chain is
\begin{equation}\label{eq:symm_ham_pca_stationary_measure}
\pi(x) 
    = \frac{Z_{x}}{Z} 
    \coloneq  \frac{\sum_{w} e^{ -H(x, w) }}{\sum_{x, w}e^{ - H(x, w) }}
\end{equation}
(where the numerator of the right hand side defines \( Z_{x} \) and the denominator defines \( Z \)).
To prove this, it is enough to show that the detailed balance condition
\( \pi(x)\operatorname{P}(x, w) = \pi(w) \operatorname{P} (w, x) \) holds (see, e.g. \cite{haggstrom2002finite}).
For the transition matrix \eqref{eq:double_hamiltonian_pca_general_form} we immediately have
\begin{equation}\label{pca_detailed_balance}
\pi(x)\operatorname{P}(x, w) 
    = \frac{Z_{x}}{Z} \frac{e^{ -H(x, w) }}{Z_{x}}
    = \frac{e^{ -H(x, w) }}{Z}
    = \frac{e^{ -H(w, x) }}{Z}
    = \frac{Z_{w}}{Z} \frac{e^{ -H(w, x) }}{Z_{w}}  
    = \pi(w) \operatorname{P}(w, x).
\end{equation}

\textbf{The \emph{lazy} PCA.}
As already mentioned, if we simply ``lift'' the Hamiltonian of the posterior distribution and update the current state of the chain according to the transition matrix defined in
\eqref{eq:double_Hamiltonian_PCA}, the stationary measure of the chain is not the posterior distribution.
To counter this issue, we modify the posterior Hamiltonian by adding an inertial term that puts a constraint on the freedom with which each component is updated at each step of the chain. We remark, however, that, conditionally on the current state, each component of the chain is updated independently of all other components at each step.

In particular, we use a pair Hamiltonian of type
\begin{equation}
    \tilde{H}(x, w) = H(x, w) + q \lVert x - w \rVert  
\end{equation}
where \( q \) is a positive parameter and \( \lVert \cdot  \rVert  \) is a suitable ``norm''.
We take \( q \lVert x - w \rVert = \sum_{i} q \lvert x_{i} - w_{i} \rvert^{p}  \),
where we use the convention \( 0^{0} = 0 \). 
In general, a Markov Chain of this type does not have the Gibbs distribution as a stationary measure. However, if the inertia of the system is big enough, the stationary measure of the chain is close to the Gibbs measure.
This argument has been made precise in the discrete context of the Ising model
in \cite{dai2012sampling, procacci2016probabilistic,apollonio2019criticality}.

Note that, if \( H(x, w) = \sum_{i=1}^{n}f_{i}(x, w_{i})\), then
\begin{equation}
\tilde{H}(x, w) = \sum_{i=1}^{n}f_{i}(x, w_{i}) + \sum_{i = 1}^{n}q \lvert x_{i} - w_{i} \rvert^{p} = \sum_{i=1}^{n}\tilde{f}_{i}(x, w_{i})
\end{equation}
and, hence, the Markov chain with transition probabilities
\begin{equation}
\operatorname{P}(x, w) = \frac{e^{- \beta \tilde{H} (x, w) }}{\sum_{w} e^{- \beta \tilde{H} (x, w) }} \propto \prod_{i = 1}^{n} e^{ -\beta \tilde{f}_{i}(x, w_{i}) }
\end{equation}
defines, again, a PCA in the sense of \eqref{eq:general_pca_definition}.

\textbf{The PCA for the posterior distribution.}
In this paper, we want to apply the \emph{lazy} PCA approach to sample from the posterior distribution 
\eqref{eq:gibbs_posterior}. We assume that the posterior Hamiltonian can be written in the form
\( H_{g}(x) = \sum_{i = 1}^{n} f_{i}(x, x_{i}) \). Then to define the lazy PCA we first lift \( H_{g}(x) \) to \( H_{g}(x, w) \) as in \eqref{eq:double_hamiltonian_pca_general_form} and then add to it the term \( q \lVert x - v \rVert  \) to obtain the posterior Hamiltonian
\begin{equation}
\tilde{H}_{g} = H_{g}(x, w) + q \lVert x - v \rVert.
\end{equation}
Then, we can define the transition probabilities as
\begin{equation}\label{eq:posterior_pca_transition_matrix}
\begin{aligned}
\operatorname{\tilde{P}}(x, w) 
    & = \frac{e^{ - \beta \tilde{H}_{g}(x, w) }}{\sum_{w} e^{ - \beta \tilde{H}_{g}(x, w) }} 
        = \frac{e^{ -\beta \sum_{i} f_{i}(x, w_{i}) + q \lvert x_{i} - w_{i} \rvert^{p}  }}{\sum_{w} e^{ -\beta \sum_{i} f_{i}(x, w_{i}) + q \lvert x_{i} - w_{i} \rvert^{p} } } \\
    & = \prod_{i = 1}^{n} \frac{e^{ - \beta \left[ f_{i}(x, w_{i}) + q\lvert x_{i} - w_{i} \rvert^{p}  \right] }}{\sum_{s \in S} e^{ - \beta \left[ f_{i}(x, s) + q\lvert x_{i} - s \rvert^{p}  \right] }}.
\end{aligned}
\end{equation}

The shape of \( f_{i} \) that must be plugged in \eqref{eq:posterior_pca_transition_matrix} depends on the particular choice of the prior distribution. In the cases we consider in this paper (see \Cref{sec:experimental_results}), $f_{i}$ is such that \( \tilde{H}_{g}(x, w) = \tilde{H}_{g}(w, x) \). Then, arguing as for \eqref{eq:symm_ham_pca_stationary_measure}, it follows that the previous transition matrix defines a Markov Chain with stationary measure
\begin{equation}
\tilde{\pi} = \frac{\sum_{w} \operatorname{\tilde{P}}(x, w)}{\sum_{x, w}\operatorname{\tilde{P}}(x, w)}.
\end{equation}

\textbf{Distinctive features of PCA.}
The primary advantage of the PCA approach to inversion is that the update rule operates completely in parallel. This means that, in principle, the computations required to update a single component of the Markov chain could be performed by a dedicated computing unit, such as a GPU core. 

It's important to note that this type of parallelism arises not from a specialized implementation of the algorithm, but from an intrinsic feature of the algorithm itself. As a result, the computational efficiency of PCA is likely to benefit ``for free'' from the technological advancements occurring in parallel computing architectures across various fields.

The counterpart to the parallel evolution is that the stationary distribution of the chain does not coincide with the target posterior distribution, but it is only an approximation. The goodness of the approximation depends on the parameter \( q \) and improves as \( q \) grows larger. 

Currently, only asymptotic results are available, showing that the stationary measure of the PCA converges to the Gibbs measure for the Ising model, which can be viewed as a case of pure black-and-white images in image restoration, as \( q \to \infty \). However, some numerical results suggest that PCA can provide quite accurate estimates for finite values of \( q \) \cite{d2021parallel,scoppola2022shaken}, and the findings presented here appear to support this observation.


\section{The Image Model}\label{sec:image_model}

In this paper, we consider gray-scale images. However, our approach could be extended to RGB images by treating each color layer separately.
Each image is encoded as a square matrix \( \Lambda \) where
each entry represents the \emph{luminance} of a pixel. 
Pixels take value in the range \( [0, 1] \) where
\( 0 \) represents pure black and \( 1 \) pure white. However, the values that can 
be taken by each pixel are not arbitrary, but are \emph{quantized} due to the fact
that only a finite number of bits is used to encode the ``intensity'' of a pixel.
We denote by \( \ell \) the number of possible levels of gray.
Note that on \( \Lambda \) we impose \emph{empty} boundary conditions, that is, the pixels at the boundary of \( \Lambda \) do not interact with other pixels and stay fixed to a default value throughout the evolution of the chain.

In what follows, we choose an ordering to explicitly identify the position and the values of the pixels of an image. Note that it is possible to add an arbitrary ordering to the elements of \( \Lambda \) (for instance, the standard \emph{column-major} ordering). When we use a single index such as \( i \) or \( j \) to denote an element (pixel) of \( \Lambda \) we refer to this arbitrary \emph{linear} ordering, whereas when we use a double indexing \( (r,c) \) when we refer to the pixel at row \( r \) and column \( c \). We call this ordering \emph{cartesian}. Clearly, there is an obvious bijection \( i \leftrightarrow (r_{i}, c_{i}) \)between the linear and the cartesian ordering.
Consequently, we write \( x_{i} \) to denote the intensity of the \( i \)-th pixel or
\( x_{(r,c)} \) for explicitely denoting the intensity of pixel at row \( r \) and column \( c \).

\begin{figure}[htbp]
    \centering
    \includegraphics[width=0.25\textwidth]{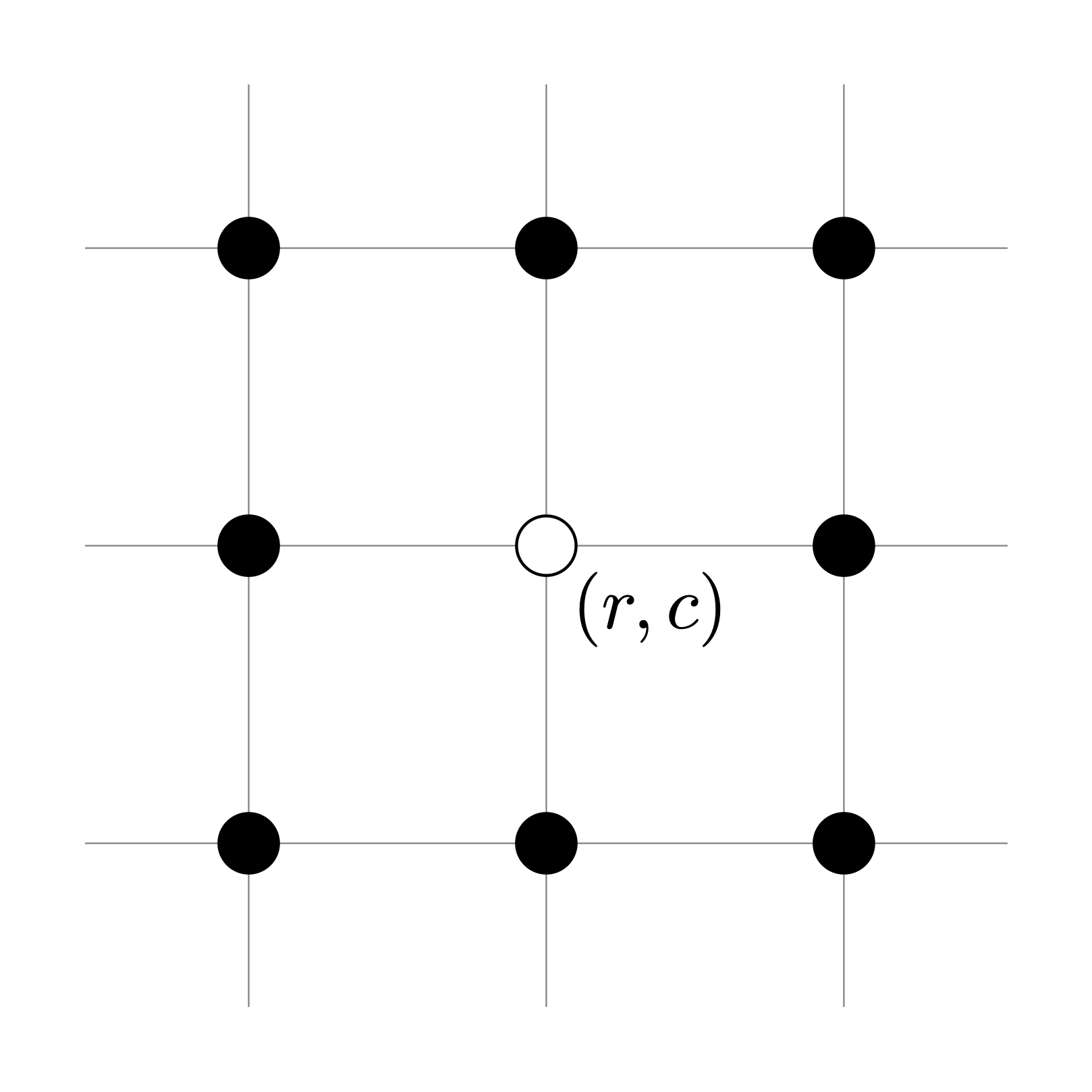}
    \caption{Neighborhood structure $\mathcal{F}(r, c)$: each pixel (center) interacts with its 8 surrounding neighbors.}
    \label{fig:neighborhood}
\end{figure}

As a prior distribution for the images taken into account, we consider the Gibbs measure with prior Hamiltonian 
\begin{equation}
    H(x) = \sum_{(r,c) \in \Lambda} f_{(r,c)}( x, x_{(r,c)})
\end{equation}
where
\begin{equation}
    f_{(r,c)} = \sum_{(p,q) \in  \mathcal{F}(r,c)} V(x_{(p,q)}, x_{(r,c)})
\end{equation}
where \( \mathcal{F}(r,c) \) is a \emph{neighborhood} of \( (r,c) \) and
\begin{equation}\label{eq:prior_pair_potential}
    V(z, w) = \begin{cases}
        -J & \text{ if } z = w \\
        +J & \text{otherwise}
    \end{cases}
\end{equation}
with \( J > 0 \). F the neighborhood \( \mathcal{F} \) we take
\begin{equation}\label{eq:neighborhood}
\mathcal{F}(r, c) = \left\{ (p, q) : \max\{ \lvert r-p \rvert, \lvert c - q \rvert   \} = 1 \right\}
\end{equation}
that is, the neighbors of \( (r, c) \) are the 8 sites of \( \Lambda \) ``surrounding'' \( (r, c) \) (see \Cref{fig:neighborhood}). Note that the number of neighbors of sites next to the boundary of the image is less than that in the ``center''.

As mentioned in \Cref{sec:background}, the Gibbs measure associated with this Hamiltonian
can be written as
\begin{equation}\label{eq:mrf_prior}
    \pi_{\mathrm{prior}}(x)  \propto e^{ - \beta H(x) } 
        = \prod_{i \in \Lambda} \exp \left\{  - \beta f_{i}(x, x_{i})  \right\}
        = \prod_{i \in  \Lambda} \exp \left\{ \beta \sum_{j \sim i} J \cdot  (2 \cdot {\mathbbm{1}}_{\left\{ x_{i} = x_{j} \right\}} - 1) \right\} 
\end{equation}
where by \( j \sim i \) we mean the sites of $\Lambda$ belonging to \( \mathcal{F}(i) \) and 
\( J 
(2 \cdot {\mathbbm{1}}_{\left\{ x_{i} = x_{j} \right\}} - 1) \) is an alternative way of writing \eqref{eq:prior_pair_potential}.
This prior probability measure favors configurations (images) where pixels are aligned with their neighbors. This tendency arises from the attractive nature of the potential defined in equation \eqref{eq:prior_pair_potential}. The use of the "double minus sign"—with one appearing in front of the Hamiltonian in the exponent of the probability measure and the other in front of the \( J \) that tunes the strength of the interaction—is common in statistical mechanics. This notation emphasizes that the configurations that occur most frequently are those that minimize the energy of the system.

Note that 
\begin{equation*}
\exp\left\{ \beta J \cdot (2 \cdot \mathbbm{1} - 1) \right\} = 
    \exp\left\{ \beta J \cdot (2 \cdot \mathbbm{1}) \right\} \cdot \exp \left\{ - \beta J \right\}
\end{equation*}
As the constant term \( \exp \left\{ - \beta J \right\} \) appears both at the numerator and the denominator of \( \pi_{\mathrm{prior}}\), it can safely be neglected.

As far as the degraded image is concerned, in this paper, we only consider the simple case of Gaussian white noise, that is
the functions \( \phi \) and \( \odot \) are,
respectively, the identity and \emph{addition}.
This means that we consider \( X - G \) (that is the difference
between the observed and the original value of a pixel)
to be normally distributed, i.e,
\begin{equation}
\pi(x,g) \propto e^{ - \frac{1}{2\sigma^{2}} \lVert g - x \rVert _{2}^{2} }
\end{equation}

Consequently, the posterior Hamiltonian can be written as
\begin{equation}\label{eq:denoising_posterior_Hamiltonian}
    H_{g} = H(x) + \frac{1}{\beta} \frac{1}{2 \sigma^{2}} \sum_{i \in  \Lambda} (g_{i} - x_{i})^{2}
\end{equation}
and the posterior distribution as
\begin{equation}\label{eq:mrf_gauss_noise_posterior_distribution}
    \pi^{\beta}_{g}(x) \coloneq \pi^{\beta}(x | g) 
        \propto e^{ - \beta H_{g}(x) }
        = \prod_{i \in  \Lambda} \exp \left\{
            \sum_{j \sim i} \beta J \cdot 2 (\mathbbm{1}_{\{ x_{i} = x_{j} \}})
              - \frac{1}{2\sigma^{2}} (g_{i} - x_{i})^{2}
        \right\}.
\end{equation}

Note that the posterior distribution is characterized by a probabilistic price to pay to have values for the pixels that are different from the observed ones.

\section{Retrieval Algorithms}\label{sec:retrieval_algorithms}

To test our PCA approach to retrieval, we compare it to the standard Gibbs Sampler algorithm. 

\subsection{Retrieval via Gibbs Sampler}
As outlined in the Introduction, the Gibbs Sampler updates the pixel at site \( i \) according to the conditional posterior probability as
\begin{equation}
    \operatorname{\mathbb{P}}\left( X_{i} = s | g, \{ X_{j} = x_{j}, j \neq  i \} \right)
        = \pi_{g}(X_{i} = s | \{ X_{j} = x_{j}, j \neq i \})
\end{equation}
which can be written as
\begin{equation}\label{eq:gibbs_sampler_elementary_step}
 \pi(X_{i} = s | g, \{X_{j} = x_{j}, j \neq  i\}) 
 = \frac{\exp 
            \left\{\left( \sum_{j \sim i} \beta J \cdot 2 (\mathbbm{1}_{\{ s = x_{j} \}}) \right)
              - \frac{1}{2\sigma^{2}} (g_{i} - s)^{2} \right\}
        }
        {
            \sum_{s \in S}
                \exp 
            \left\{\left( \sum_{j \sim i} \beta J \cdot 2 (\mathbbm{1}_{\{ s = x_{j} \}}) \right)
              - \frac{1}{2\sigma^{2}} (g_{i} - s)^{2} \right\}
        }
\end{equation}
where \( S \) is the set of possible gray levels.

The retrieval procedure performs a sequence of \emph{sweeps} where
each sweep consists of a sequence of steps, each as in \eqref{eq:gibbs_sampler_elementary_step},
performed on the sites of \( \Lambda \) taken in a predetermined order (for instance, the column-major one).

The parameter \( \beta \), which is the inverse of the temperature, can, in principle, be varied during the retrieval process. The so-called \emph{simulated annealing} is performed by increasing \( \beta \) throughout the evolution of the chain according to a suitable schedule.

\subsection{Retrieval via PCA}

As discussed in \Cref{sec:probabilistic_cellular_automata}, a PCA is a Markov chain
where transition probabilities are of the form of \eqref{eq:double_Hamiltonian_PCA}
and are defined in terms of a suitable ``pair Hamiltonian''. 
The key feature of this pair Hamiltonian is that it can be written in terms of a sum over the sites of \( \Lambda \).
In an inversion problem such as the one we consider in this paper, we want the pair Hamiltonian to be a ``lifted'' version of the posterior Hamiltonian \eqref{eq:denoising_posterior_Hamiltonian} as explained at the beginning of \Cref{sec:probabilistic_cellular_automata}. 
Further, to put forward our \emph{lazy} PCA approach, we add the posterior pair Hamiltonian
an inertial term which is proportional to a suitable \emph{distance} between the configuration corresponding to the two arguments of the pair Hamiltonian.
Here we consider the following \emph{posterior pair Hamiltonian}
\begin{equation}\label{eq:denoising_pair_posterior_Hamiltonian}
\begin{aligned}
    \tilde{H}_{g}(x, w) 
        & = H_{g}(x, w) + q \lVert x - w \rVert _{0} \\
        & =  \sum_{i  \in \Lambda} \left( - \sum_{j \sim i} J \cdot 2(\mathbbm{1}_{\{ x_{i} = x_{j} \}}) \right)
             + \frac{1}{\beta} \frac{1}{2 \sigma^{2}} \sum_{i \in  \Lambda} (g_{i} - x_{i})^{2}
             + q \sum_{i} \mathbbm{1}_{\{ x_{i} \neq  w_{i} \}}
\end{aligned}
\end{equation}
where \( q \) is a positive parameter.

With these definitions, the retrieval procedure consists of a sequence of steps where
at each step the probability of going from \( x \) to \( w \) is
\begin{equation}
\begin{aligned}
    \operatorname{P}(x, w) 
        & \propto e^{ -\beta \tilde{H}_{g}(x, w) } \\ 
        & = \prod_{i \in \Lambda} \frac{\exp 
                    \left\{\left( \sum_{j \sim i} \beta J \cdot 2 (\mathbbm{1}_{\{ s = x_{j} \}}) \right)
                    - \frac{1}{2\sigma^{2}} (g_{i} - s)^{2} - \beta q \mathbbm{1}_{\{ s \neq x_{j} \}} \right\}
                }
                {
                    \sum_{s \in S}
                        \exp 
                    \left\{\left( \sum_{j \sim i} \beta J \cdot 2 (\mathbbm{1}_{\{ s = x_{j} \}}) \right)
                    - \frac{1}{2\sigma^{2}} (g_{i} - s)^{2} - \beta q \mathbbm{1}_{\{ s \neq x_{j} \}} \right\}
                }
\end{aligned}
\end{equation}
where, again, \( S \) is the set of possible gray levels.
Note that, at each step, all pixels are potentially updated.

The effect of the additional inertial terms \(-\beta q \mathbbm{1}_{\{ s \neq x_{j} \}} \) in the exponent of each factor of the transition probabilities is to add a probabilistic price every time the value of a pixel changes.
Since the potential \eqref{eq:prior_pair_potential} we are taking into account
is only concerned with whether two pixels in the same neighborhood have the same value or not,
the \( L_{0} \) \emph{norm} we consider for the inertial term seems to be appropriate. Note that this is not a restriction and other norms such as the \( L_{1} \) or  \( L_{2} \) can be considered. These norms are,
likely, more suitable when $L_{1}$ or \( L_{2} \) norms appear in the prior. Then the probabilistic price to update the value of a pixel is proportional to the \emph{size} of the change.


\section{Experimental Results}\label{sec:experimental_results}

To test our algorithm, we consider the restoration of noisy images.
In particular, most of the images we consider are, ideally, samples of a Markov Random Field (MRF) to which we add Gaussian Noise. Note that a MRF is a probability measure that can be specified via a Gibbs measure $\pi_{\mathrm{prior}}(x) = \frac{e^{- \beta H(x)}}{Z}$ (see \cite{geman1984stochastic} and \cite{bremaud2020markov} for more details).

This is the same setup as the simpler cases considered in \cite{geman1984stochastic}, which we find suitable for an initial assessment of the PCA approach to inverse problems.

To generate the MRF, we start a chain from a set of pixels randomly chosen from $\ell$ gray levels (equally spaced from 0, pure black to 1, pure white) and let it evolve according to the Gibbs Sampler algorithm with Gibbs measure
\begin{equation}
    \pi_{\mathrm{prior}} = \frac{e^{-\beta H(x)}}{Z}
\end{equation}
with \( H \) as in \eqref{eq:mrf_prior} and \( J = \frac{1}{3} \). We perform a manual adjustment of \( \beta \) during the evolution of the chain to obtain ``not too noisy'' original images.
Then the degraded images are obtained by adding independent normally distributed values (with mean zero and standard deviation \( \sigma \)) to each pixel and rounding the obtained result to the nearest gray level.

Both algorithms are run for a fixed number of steps (sweeps in the case of Gibbs Sampler), and we consider the last sample as the retrieved image. The aim is to perform a MAP estimate of the original image. 
We consider MRF images with \( 5 \), \( 9 \) and 33 gray levels and set the standard deviation of the noise to \( 0.25 \) for the images with \( 5 \) gray levels, to \( 0.2 \) for the image with \( 9 \) gray levels and to \( 0.1 \) for the image with \( 33 \) gray levels.
Note that, in all cases, the standard deviation of the noise is larger than the gap between two consecutive levels. These images have \( 256 \times 256 \) or \( 512 \times 512 \) pixels.

We also consider two simple non-MRF black and white images taken from the \href{https://www.imageprocessingplace.com/DIP-3E/dip3e_book_images_downloads.htm}{Image Processing Place} database. These images are chosen since they present uniform regions without sharp edges and are therefore well adapted to be treated with the simple prior model we use for this initial work on Bayesian inversion via PCA.

For both algorithms and all test cases, we perform 1000 steps with the following heuristics cooling schedule for \( \beta \): we start with \( \beta = 1.25 \) and increase it by \( 0.25 \) every 250 steps. We set \( \sigma \) equal to the value of the added noise. 
Note that the PCA algorithm also depends on the inertial parameter \( q \).
For the PCA, we test \( 14 \) values of \( q \) in the interval \( [0.06, 0.71] \), with a step of \( 0.05 \). A discussion on the dependence of the performance of the algorithm on the value of \( q \) is given below in \Cref{sec:q-robustness}.

\begin{figure}[htbp]
    \centering
    \begin{subfigure}[t]{0.2\textwidth}
        \centering
        \includegraphics[width=\textwidth]{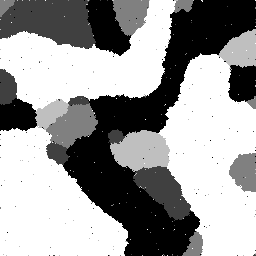}
        \caption{Original image.}
        \label{fig:subfig1}
    \end{subfigure}
    \qquad
    \begin{subfigure}[t]{0.2\textwidth}
        \centering
        \includegraphics[width=\textwidth]{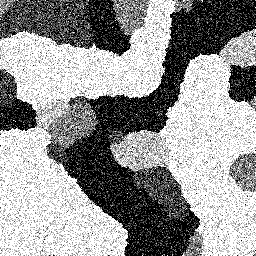}
        \caption{Noisy image.}
        \label{fig:subfig2}
    \end{subfigure}
    \qquad
    \begin{subfigure}[t]{0.2\textwidth}
        \centering
        \includegraphics[width=\textwidth]{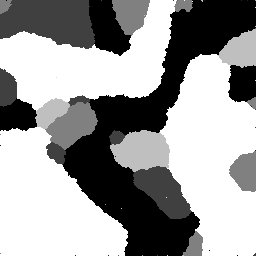}
        \caption{Denoised with Gibbs Sampler.}
        \label{fig:subfig3}
    \end{subfigure}
    \qquad
    \begin{subfigure}[t]{0.2\textwidth}
        \centering
        \includegraphics[width=\textwidth]{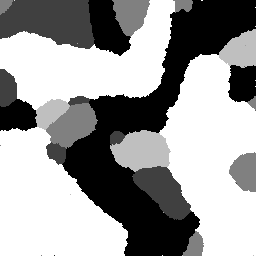}
        \caption{Denoised with PCA.}
        \label{fig:subfig4}
    \end{subfigure}
    \caption{Image mrf\_n256\_l5\_i00 - 5 levels of gray; the standard deviation of the white noise is $\sigma = 0.25$. (First example)}
    \label{fig:l5-1}
\end{figure}

\begin{figure}[htbp]
    \centering
    \begin{subfigure}[t]{0.2\textwidth}
        \centering
        \includegraphics[width=\textwidth]{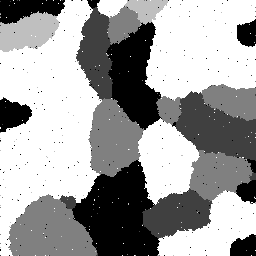}
        \caption{Original image.}
        \label{fig:subfig1}
    \end{subfigure}
    \qquad
    \begin{subfigure}[t]{0.2\textwidth}
        \centering
        \includegraphics[width=\textwidth]{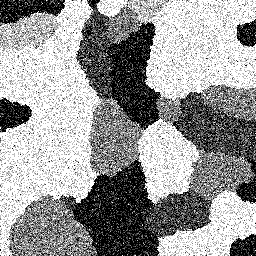}
        \caption{Noisy image.}
        \label{fig:subfig2}
    \end{subfigure}
    \qquad
    \begin{subfigure}[t]{0.2\textwidth}
        \centering
        \includegraphics[width=\textwidth]{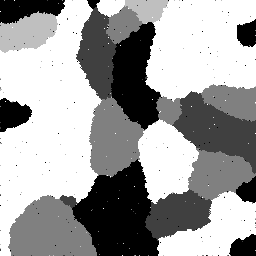}
        \caption{Denoised with Gibbs Sampler.}
        \label{fig:subfig3}
    \end{subfigure}
    \qquad
    \begin{subfigure}[t]{0.2\textwidth}
        \centering
        \includegraphics[width=\textwidth]{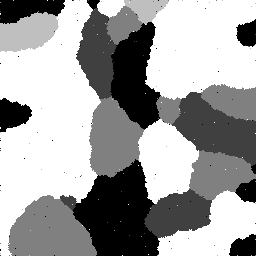}
        \caption{Denoised with PCA.}
        \label{fig:subfig4}
    \end{subfigure}
    \caption{mrf\_n256\_l5\_i002 - 5 levels of gray, the standard deviation of the white noise is $\sigma = 0.25$. (Second example)}
    \label{fig:l5-2}
\end{figure}

\begin{figure}[htbp]
    \centering
    \begin{subfigure}[t]{0.2\textwidth}
        \centering
        \includegraphics[width=\textwidth]{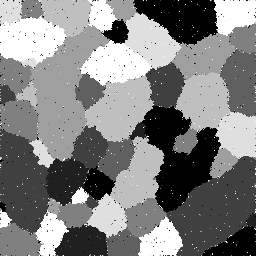}
        \caption{Original image.}
        \label{fig:subfig1}
    \end{subfigure}
    \qquad
    \begin{subfigure}[t]{0.2\textwidth}
        \centering
        \includegraphics[width=\textwidth]{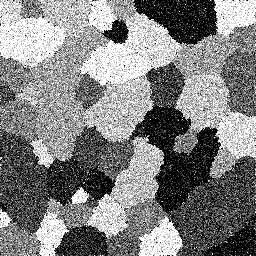}
        \caption{Noisy image.}
        \label{fig:subfig2}
    \end{subfigure}
    \qquad
    \begin{subfigure}[t]{0.2\textwidth}
        \centering
        \includegraphics[width=\textwidth]{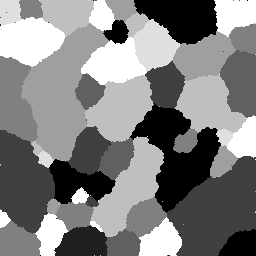}
        \caption{Denoised with Gibbs Sampler.}
        \label{fig:subfig3}
    \end{subfigure}
    \qquad
    \begin{subfigure}[t]{0.2\textwidth}
        \centering
        \includegraphics[width=\textwidth]{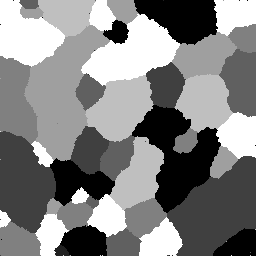}
        \caption{Denoised with PCA.}
        \label{fig:subfig4}
    \end{subfigure}
    \caption{mrf\_n256\_l9\_i001 - 9 levels of gray; the standard deviation of the white noise is $\sigma = 0.20$.}
    \label{fig:l9-1}
\end{figure}

\begin{figure}[htbp]
    \centering
    \begin{subfigure}[t]{0.2\textwidth}
        \centering
        \includegraphics[width=\textwidth]{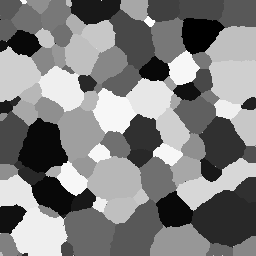}
        \caption{Original image.}
        \label{fig:subfig1}
    \end{subfigure}
    \qquad
    \begin{subfigure}[t]{0.2\textwidth}
        \centering
        \includegraphics[width=\textwidth]{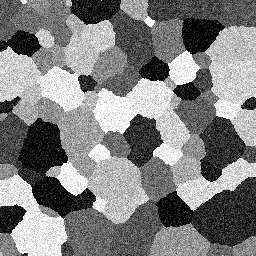}
        \caption{Noisy image.}
        \label{fig:subfig2}
    \end{subfigure}
    \qquad
    \begin{subfigure}[t]{0.2\textwidth}
        \centering
        \includegraphics[width=\textwidth]{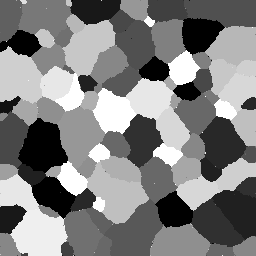}
        \caption{Denoised with Gibbs Sampler.}
        \label{fig:subfig3}
    \end{subfigure}
    \qquad
    \begin{subfigure}[t]{0.2\textwidth}
        \centering
        \includegraphics[width=\textwidth]{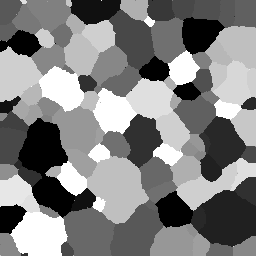}
        \caption{Denoised with PCA.}
        \label{fig:subfig4}
    \end{subfigure}
    \caption{mrf\_n256\_l33\_i001 - 33 levels of gray; the standard deviation of the white noise is $\sigma = 0.10$.}
    \label{fig:l33-1}
\end{figure}

\begin{figure}[htbp]
    \centering
    \begin{subfigure}[t]{0.2\textwidth}
        \centering
        \includegraphics[width=\textwidth]{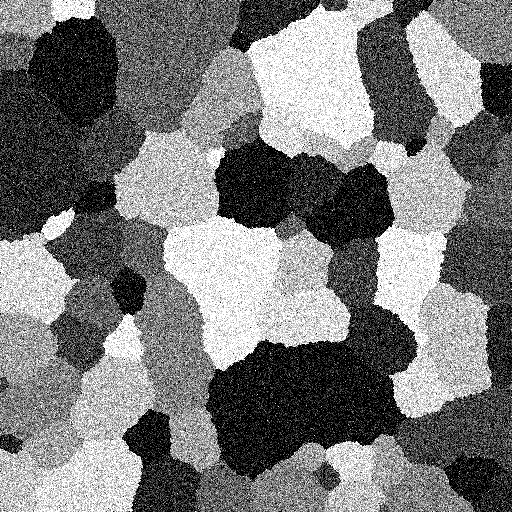}
        \caption{Original image.}
        \label{fig:subfig1}
    \end{subfigure}
    \qquad
    \begin{subfigure}[t]{0.2\textwidth}
        \centering
        \includegraphics[width=\textwidth]{mrf_n512_l9_i001-s20.png}
        \caption{Noisy image.}
        \label{fig:subfig2}
    \end{subfigure}
    \qquad
    \begin{subfigure}[t]{0.2\textwidth}
        \centering
        \includegraphics[width=\textwidth]{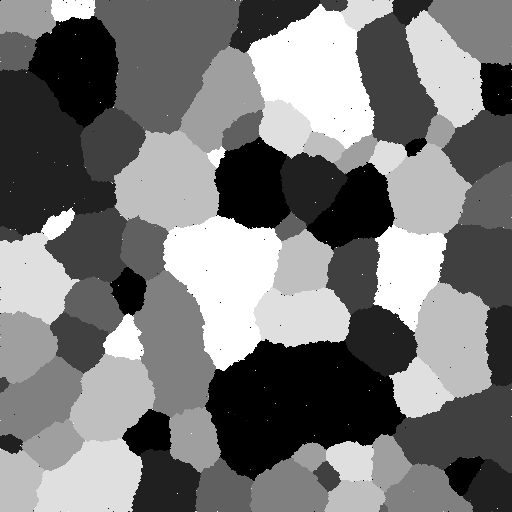}
        \caption{Denoised with Gibbs Sampler.}
        \label{fig:subfig3}
    \end{subfigure}
    \qquad
    \begin{subfigure}[t]{0.2\textwidth}
        \centering
        \includegraphics[width=\textwidth]{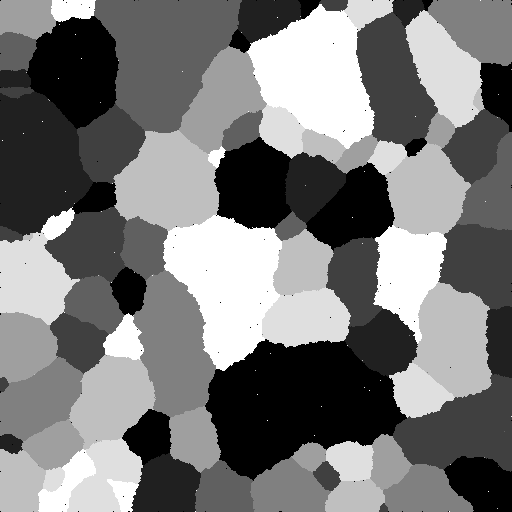}
        \caption{Denoised with PCA.}
        \label{fig:subfig4}
    \end{subfigure}
    \caption{mrf\_n512\_l9\_i001 - 9 levels of gray; the standard deviation of the white noise is $\sigma = 0.20$.}
    \label{fig:l9-512-1}
\end{figure}

\begin{figure}[htbp]
    \centering
    \begin{subfigure}[t]{0.2\textwidth}
        \centering
        \includegraphics[width=\textwidth]{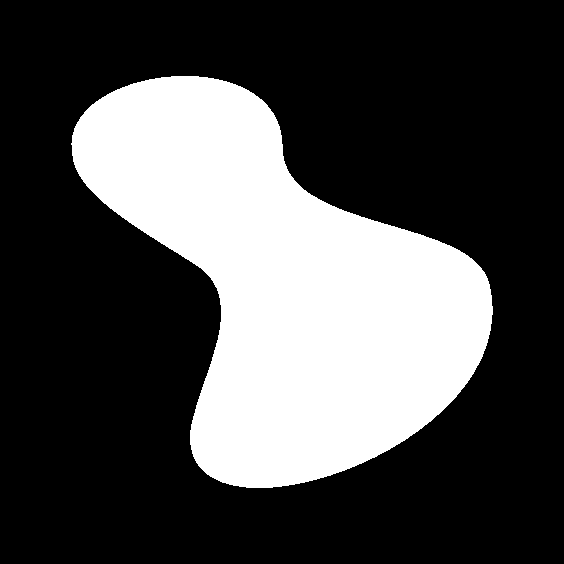}
        \caption{Original image.}
        \label{fig:subfig1}
    \end{subfigure}
    \qquad
    \begin{subfigure}[t]{0.2\textwidth}
        \centering
        \includegraphics[width=\textwidth]{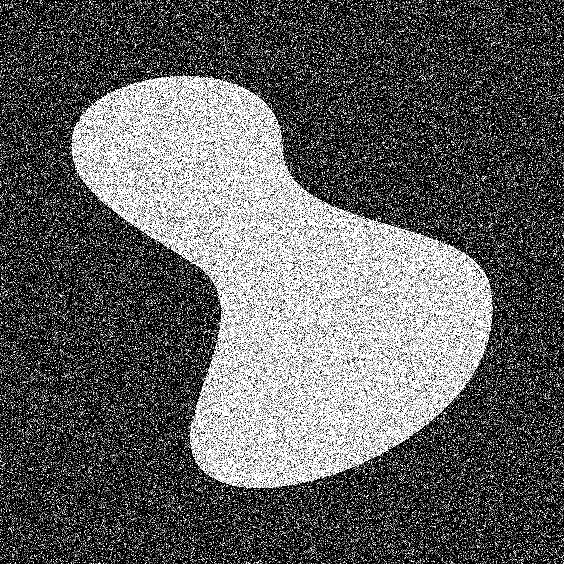}
        \caption{Noisy image.}
        \label{fig:subfig2}
    \end{subfigure}
    \qquad
    \begin{subfigure}[t]{0.2\textwidth}
        \centering
        \includegraphics[width=\textwidth]{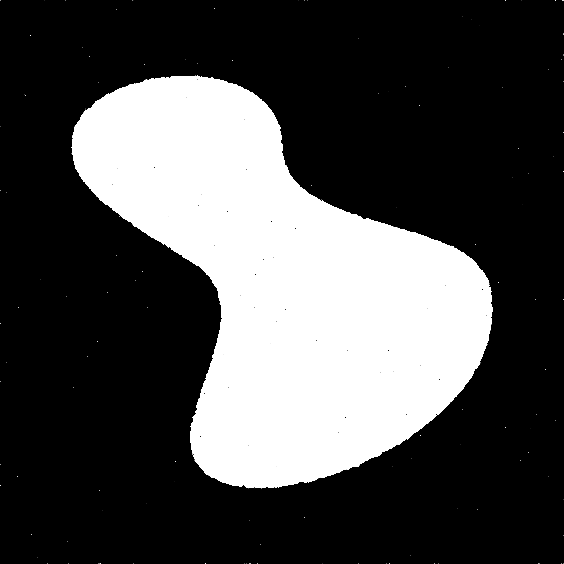}
        \caption{Denoised with PCA.}
        \label{fig:subfig3}
    \end{subfigure}
    \qquad
     \begin{subfigure}[t]{0.2\textwidth}
        \centering
        \includegraphics[width=\textwidth]{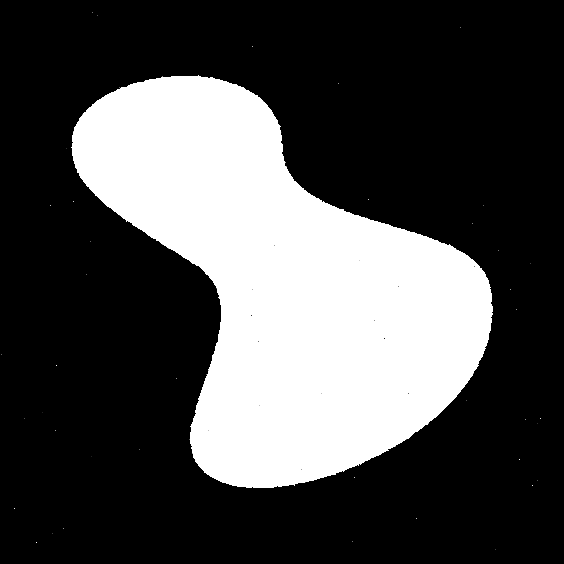}
        \caption{Denoised with PCA.}
        \label{fig:subfig3}
    \end{subfigure}\\
    \begin{subfigure}[t]{0.2\textwidth}
        \centering
        \includegraphics[width=\textwidth]{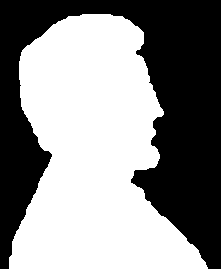}
        \caption{Original image.}
        \label{fig:subfig1}
    \end{subfigure}
    \qquad
    \begin{subfigure}[t]{0.2\textwidth}
        \centering
        \includegraphics[width=\textwidth]{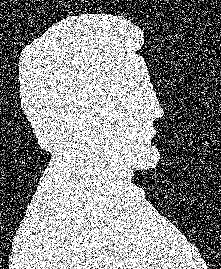}
        \caption{Noisy image.}
        \label{fig:subfig2}
    \end{subfigure}
    \qquad
    \begin{subfigure}[t]{0.2\textwidth}
        \centering
        \includegraphics[width=\textwidth]{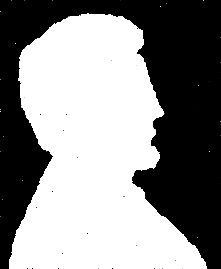}
        \caption{Denoised with PCA.}
        \label{fig:subfig3}
    \end{subfigure}
    \begin{subfigure}[t]{0.2\textwidth}
        \centering
        \includegraphics[width=\textwidth]{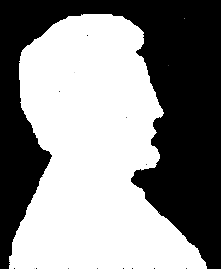}
        \caption{Denoised with PCA.}
        \label{fig:subfig3}
    \end{subfigure}
    \caption{Two examples (``Blob'' above and ``Lincoln'' below) of non-randomly generated images denoised with the PCA. For both images, the standard deviation of the noise is \( 0.5 \). The denoising has been performed with 1000 sweeps/steps with \( \beta \) kept fixed at the value \( 2.0 \). For the PCA the value of \( q \) has been set to \( 0.51 \).}
    \label{fig:real-images}
\end{figure}

Examples of the retrieval (denoising) for MRF images are provided in \Crefrange{fig:l5-1}{fig:l9-512-1}, whereas examples of results obtained on the non-MRF images are illustrated in \Cref{fig:real-images}.

Both algorithms have been evaluated in terms of the Structural similarity index measure (SSIM) and the Peak signal-to-noise ratio (PSNR).

If \( x \) is the original \( m \times n \) image and \(\tilde{x} \) is its noisy version, the 
PSNR of \( \tilde{x} \) with respect to \( x \) is
\begin{equation}
\operatorname{PSNR}(\tilde{x}) = 20 \log_{10} \left( \frac{\max x}{\sqrt{\operatorname{MSE}(x, \tilde{x})}} \right)
\end{equation}
where \( \max x \) is the maximum pixel value of the original image, 
\begin{equation}
\operatorname{MSE}(x, \tilde{x}) = \frac{1}{m \cdot n} \sum_{i = 1}^{m} \sum_{j = 1}^{n} \left[ x_{(i,j)} - \tilde{x}_{(i, j)} \right]^{2}
\end{equation}
is the Mean Square Error of \( x \) and \( \tilde{x} \), and \( x_{(i,j)}, \, \tilde{x}_{(i,j)} \) are the luminances of the pixel with coordinates \( (i, j) \).

SSIM has been introduced in \cite{wang2004image} to measure the \emph{perceived similarity} between two images \( x \) and \( y \) and
it is defined as
\begin{equation}
\operatorname{SSIM}(x, y) 
    = \frac{(2\mu_{x} \mu_{y} + c_{1}) (2 \sigma_{xy} + c_{2})}
           {(\mu_{x}^{2} + \mu_y^{2} + c_{1})(\sigma_{x}^{2} + \sigma_y^{2} + c_{2})}
\end{equation}
where \( \mu_{x}, \mu_{y} \) are the sample means of \( x \) and \( y \), \( \sigma_{x}, \sigma_{y} \) are their sample standard deviations and \( \sigma_{xy} \) is the sample covariance.
Constants \( c_{1} \) and \( c_{2} \) depend, essentially on the number of possible gray levels
and are meant to stabilize the results in the case of small denominators.

The obtained results are provided in
\Cref{tab:algo_comparison} for the MRF images and in \Cref{fig:real-images} for the non-MRF images.
In most cases, the average values of both SSIM and PSNR of the PCA-restored images is always higher than that of the images restored using the Gibbs sampler.


\begin{table}[ht]
\centering
\small
\setlength{\tabcolsep}{4pt}
\begin{tabular}{@{}
                l
                c
                c
                c
                c
                c
                S[table-format=1.4,round-mode=places,round-precision=4]
                S[table-format=2.4,round-mode=places,round-precision=4]
                S[table-format=1.4,round-mode=places,round-precision=4]
                S[table-format=2.4,round-mode=places,round-precision=4]
                @{}}
\toprule
\textbf{image id} &
{$N$} &
{$\sigma$} &
{$\ell$} &
{\textbf{algo}} &
{$q$} &
\multicolumn{2}{c}{\textbf{Average}} &
\multicolumn{2}{c}{\textbf{Best}} \\
\cmidrule(lr){7-8} \cmidrule(lr){9-10}
&
&
&
&
&
&
{\textbf{SSIM}} &
{\textbf{PSNR}} &
{\textbf{SSIM}} &
{\textbf{PSNR}} \\
\midrule
\multirow{2}{*}{mrf\_n256\_l5\_i001}
& \multirow{2}{*}{256}
& \multirow{2}{*}{0.25}
& \multirow{2}{*}{5}
& GS
& --
& 0.845328 +- 0.0000
& 24.832526 +- 0.0000
& 0.852979
& 25.205007 \\
&
&
&
&
PCA
& $0.56$
& 0.851266 +- 0.0000
& 25.140595 +- 0.0000
& 0.855304
& 25.381551 \\
\midrule
\multirow{2}{*}{mrf\_n256\_l5\_i002}
& \multirow{2}{*}{256}
& \multirow{2}{*}{0.25}
& \multirow{2}{*}{5}
& GS
& --
& 0.772941 +- 0.0000
& 22.917390 +- 0.0000
& 0.779175
& 23.029294 \\
&
&
&
&
PCA
& $0.56$
& 0.776616 +- 0.0000
& 23.140142 +- 0.0000
& 0.778275
& 23.218906 \\
\midrule
\multirow{2}{*}{mrf\_n256\_l9\_i001}
& \multirow{2}{*}{256}
& \multirow{2}{*}{0.20}
& \multirow{2}{*}{9}
& GS
& --
& 0.697332 +- 0.0000
& 22.344301 +- 0.0000
& 0.703129
& 22.537597 \\
&
&
&
&
PCA
& $0.11$
& 0.701505 +- 0.0000
& 22.498807 +- 0.0000
& 0.707730
& 22.631849 \\
\midrule
\multirow{2}{*}{mrf\_n256\_l33\_i001}
& \multirow{2}{*}{256}
& \multirow{2}{*}{0.10}
& \multirow{2}{*}{33}
& GS
& --
& 0.906450 +- 0.0000
& 30.607542 +- 0.0000
& 0.913825
& 31.113281 \\
&
&
&
&
PCA
& $0.31$
& 0.901192 +- 0.0000
& 30.549015 +- 0.0000
& 0.904500
& 30.953039 \\
\midrule
\multirow{2}{*}{mrf\_n512\_l9\_i001}
& \multirow{2}{*}{512}
& \multirow{2}{*}{0.20}
& \multirow{2}{*}{9}
& GS
& --
& 0.857077 +- 0.0000
& 26.598590 +- 0.0000
& 0.860189
& 26.719179 \\
&
&
&
&
PCA
& $0.26$
& 0.863556 +- 0.0000
& 26.838477 +- 0.0000
& 0.867689
& 27.068381 \\
\bottomrule
\end{tabular}
\caption{Average (over 10 runs) SSIM and PSNR values and best SSIM and PSNR values for the MRF images denoised via the Gibbs Sampler (GS) and the Probabilistic Cellular Automaton (PCA). For the PCA the table presents the values obtained with the \( q \) which yielded the larges average value of SSIM.}
\label{tab:algo_comparison}
\end{table}

\begin{table}[ht]
\centering
\small
\setlength{\tabcolsep}{4pt}
\begin{tabular}{@{}
                l
                c
                c
                c
                c
                c
                S[table-format=1.4,round-mode=places,round-precision=4]
                S[table-format=2.4,round-mode=places,round-precision=4]
                S[table-format=1.4,round-mode=places,round-precision=4]
                S[table-format=2.4,round-mode=places,round-precision=4]
                @{}}
\toprule
\textbf{image id} &
{\textbf{size}} &
{$\sigma$} &
{$\ell$} &
{\textbf{algo}} &
{$q$} &
\multicolumn{2}{c}{\textbf{Average}} &
\multicolumn{2}{c}{\textbf{Best}} \\
\cmidrule(lr){7-8} \cmidrule(lr){9-10}
&
&
&
&
&
&
{\textbf{SSIM}} &
{\textbf{PSNR}} &
{\textbf{SSIM}} &
{\textbf{PSNR}} \\
\midrule
\multirow{2}{*}{Blob}
& \multirow{2}{*}{564 $\times$ 564}
& \multirow{2}{*}{0.5}
& \multirow{2}{*}{5}
& GS
& --
& 0.986787 +- 0.0000
& 29.634207 +- 0.0000
& 0.987762
& 29.892574 \\
&
&
&
&
PCA
& $0.71$
& 0.993059 +- 0.0000
& 31.171271 +- 0.0000
& 0.993778
& 31.526739 \\
\midrule
\multirow{2}{*}{Lincoln}
& \multirow{2}{*}{269 $\times$ 221}
& \multirow{2}{*}{0.5}
& \multirow{2}{*}{5}
& GS
& --
& 0.971809 +- 0.0000
& 25.119581 +- 0.0000
& 0.974493
& 25.424202 \\
&
&
&
&
PCA
& $0.46$
& 0.980930 +- 0.0000
& 26.472092 +- 0.0000
& 0.982954
& 25.853918 \\
\bottomrule
\end{tabular}
\caption{Average (over 10 runs) SSIM and PSNR values and best SSIM and PSNR values for the non-MRF images denoised via the Gibbs Sampler (GS) and the Probabilistic Cellular Automaton (PCA). For the PCA the table presents the values obtained with the \( q \) which yielded the larges average value of SSIM.}
\label{tab:algo_comparison-real-images}
\end{table}

However, the advantages of PCA are more relevant in terms of execution time (``wall clock''). The execution times for 1000 ``sweeps'' of the Gibbs Sampler and 1000 steps of the PCA are presented in \Cref{tab:time-comparison}.
In our experimental setting (see \Cref{sec:experimental_details} below), the PCA is between 200 and 300 times faster than its sequential counterpart.
We remark that the increased computational effectiveness is not due to an especially crafted implementation of the PCA algorithm (the computation kernel for the PCA and the Gibbs Sampler are the same), but is due to the intrinsic parallel nature of the algorithm, which allows us to exploit the computing resources of the GPU in a straightforward manner.

\begin{table}[h!]
\centering
\begin{tabular}{@{}ccccc@{}}
\toprule
$N$   & $\ell$ & \textbf{time GS }& \textbf{time PCA} & \textbf{speedup PCA} \\ \midrule
256 & 5      & 16.59        & 0.0776     & 213         \\
256 & 9      & 28.11        & 0.1317     & 213         \\
256 & 33     & 100.6        & 0.41       & 245         \\
512 & 9      & 108          & 0.352      & 306         \\ \bottomrule
\end{tabular}
\caption{Execution times (in seconds) for 1000 ``sweeps'' of the single spin flip Gibbs Sampler (GS) running on the CPU and 1000 steps of the PCA running on the GPU. The running times of the PCA include the load and unload of the data to and from the GPU .}
\label{tab:time-comparison}
\end{table}

\subsection{Sensitivity of the results respect to \( q \)}\label{sec:q-robustness}

A delicate feature concerning the application of the PCA algorithm to Bayesian inversion is its dependence on one additional parameter, namely the inertial parameter \( q \). Indeed, in principle, values of \( q \) that are too low yield a stationary distribution that is far from the desired posterior Gibbs distribution. On the other hand, if \( q \) is too large, the chain tends, at each step, to update a rather limited number of pixels, thus hindering the benefits coming from parallelism.

Nevertheless, in this particular application, the procedure turns out to be relatively robust with respect to the variation of the parameter \( q \) in the interval \( [0.06, 0.71] \)
In \Cref{tab:q-sensitivity-all-images} below, we give the average and best values of SSIM and PSNR obtained for several values of  \( q \) for the test cases taken into consideration.

The collected data seem to suggest that the algorithm’s "performance" may be a concave function of \( q \). If this were indeed the case, one could use an efficient strategy to find the optimal value of \( q \) provided one has access to a measure of goodness for the inversion process (for instance, the value of the posterior Hamiltonian that is reached after a certain number of steps). The additional number of times the algorithm must be run to perform the inversion for several values of \( q \) should still be more than compensated for by the large speedup one obtains using PCA instead of SSF.
Assessing the concavity property of the "goodness" of the inversion process should be the subject of a subsequent study.

{
\small
\setlength{\LTleft}{0pt}
\setlength{\LTright}{0pt}
\begin{longtable}{@{}ccccc@{}}
\caption{Performance of the PCA algorithm as a function of \( q \). For each image and for each tested value of \( q \), the table reports the mean and standard deviation of SSIM and PSNR over 10 runs, together with the best SSIM and PSNR obtained among the 10 runs. For each image, the best mean SSIM, the best mean PSNR, the best SSIM, and the best PSNR are highlighted in bold.}
\label{tab:q-sensitivity-all-images}\\
\toprule
$q$ & \textbf{SSIM (mean $\pm$ std)} & \textbf{PSNR (mean $\pm$ std)} & \textbf{SSIM (best)} & \textbf{PSNR (best)} \\
\midrule
\endfirsthead

\toprule
$q$ & \textbf{SSIM (mean $\pm$ std)} & \textbf{PSNR (mean $\pm$ std)} & \textbf{SSIM (best)} & \textbf{PSNR (best)} \\
\midrule
\endhead

\midrule
\multicolumn{5}{r}{\emph{Continued on next page}}\\
\endfoot

\bottomrule
\endlastfoot

\multicolumn{5}{@{}l}{\textbf{Image: mrf\_n256\_l5\_i001}}\\
0.06 & $0.8447 \pm 0.0024$ & $24.7787 \pm 0.1037$ & $0.8479$ & $24.9349$ \\
0.11 & $0.8448 \pm 0.0029$ & $24.7842 \pm 0.1183$ & $0.8476$ & $24.9194$ \\
0.16 & $0.8452 \pm 0.0021$ & $24.7987 \pm 0.1042$ & $0.8482$ & $24.9439$ \\
0.21 & $0.8464 \pm 0.0018$ & $24.8468 \pm 0.1116$ & $0.8491$ & $25.0077$ \\
0.26 & $0.8471 \pm 0.0018$ & $24.9271 \pm 0.0966$ & $0.8501$ & $25.0698$ \\
0.31 & $0.8489 \pm 0.0018$ & $25.0057 \pm 0.1082$ & $0.8518$ & $25.1288$ \\
0.36 & $0.8488 \pm 0.0024$ & $24.9833 \pm 0.1068$ & $0.8515$ & $25.1572$ \\
0.41 & $0.8490 \pm 0.0031$ & $25.0021 \pm 0.1437$ & $0.8525$ & $25.1913$ \\
0.46 & $0.8496 \pm 0.0018$ & $25.0827 \pm 0.0741$ & $0.8528$ & $25.2353$ \\
0.51 & $0.8502 \pm 0.0019$ & $25.0545 \pm 0.0344$ & $0.8534$ & $25.1220$ \\
0.56 & $\mathbf{0.8513 \pm 0.0022}$ & $25.1406 \pm 0.1166$ & $0.8553$ & $\mathbf{25.3816}$ \\
0.61 & $0.8509 \pm 0.0018$ & $\mathbf{25.1415 \pm 0.0826}$ & $0.8534$ & $25.2353$ \\
0.66 & $0.8511 \pm 0.0013$ & $25.1135 \pm 0.0598$ & $0.8536$ & $25.2023$ \\
0.71 & $0.8506 \pm 0.0024$ & $25.1410 \pm 0.1302$ & $\mathbf{0.8556}$ & $25.3801$ \\
\midrule

\multicolumn{5}{@{}l}{\textbf{Image: mrf\_n256\_l5\_i002}}\\
0.06 & $0.7709 \pm 0.0027$ & $22.8960 \pm 0.0712$ & $0.7752$ & $23.0218$ \\
0.11 & $0.7720 \pm 0.0022$ & $22.9283 \pm 0.0737$ & $0.7745$ & $23.0251$ \\
0.16 & $0.7736 \pm 0.0018$ & $22.9653 \pm 0.0809$ & $0.7773$ & $23.1184$ \\
0.21 & $0.7736 \pm 0.0027$ & $23.0276 \pm 0.0731$ & $0.7786$ & $23.1337$ \\
0.26 & $0.7741 \pm 0.0021$ & $23.0199 \pm 0.0534$ & $0.7764$ & $23.1158$ \\
0.31 & $0.7743 \pm 0.0021$ & $23.0155 \pm 0.0568$ & $0.7788$ & $23.0888$ \\
0.36 & $0.7750 \pm 0.0012$ & $23.0747 \pm 0.0674$ & $0.7770$ & $23.2215$ \\
0.41 & $0.7753 \pm 0.0018$ & $23.0822 \pm 0.0501$ & $0.7768$ & $23.1576$ \\
0.46 & $0.7754 \pm 0.0024$ & $23.0848 \pm 0.1077$ & $0.7789$ & $\mathbf{23.2337}$ \\
0.51 & $0.7757 \pm 0.0027$ & $23.0838 \pm 0.0893$ & $0.7780$ & $23.2250$ \\
0.56 & $\mathbf{0.7766 \pm 0.0016}$ & $23.1401 \pm 0.0534$ & $0.7783$ & $23.2189$ \\
0.61 & $0.7753 \pm 0.0017$ & $23.0971 \pm 0.0560$ & $0.7774$ & $23.1826$ \\
0.66 & $0.7758 \pm 0.0021$ & $\mathbf{23.1415 \pm 0.0499}$ & $0.7790$ & $23.2311$ \\
0.71 & $0.7764 \pm 0.0021$ & $23.1260 \pm 0.0681$ & $\mathbf{0.7801}$ & $23.2293$ \\
\midrule

\multicolumn{5}{@{}l}{\textbf{Image: mrf\_n256\_l9\_i001}}\\
0.06 & $0.6985 \pm 0.0053$ & $22.4268 \pm 0.1368$ & $0.7074$ & $22.5941$ \\
0.11 & $\mathbf{0.7015 \pm 0.0029}$ & $\mathbf{22.4988 \pm 0.0934}$ & $0.7077$ & $22.6318$ \\
0.16 & $0.6982 \pm 0.0062$ & $22.4022 \pm 0.1750$ & $0.7061$ & $22.6125$ \\
0.21 & $0.6991 \pm 0.0059$ & $22.4328 \pm 0.1471$ & $\mathbf{0.7095}$ & $22.6899$ \\
0.26 & $0.7003 \pm 0.0042$ & $22.4439 \pm 0.2121$ & $0.7046$ & $22.6681$ \\
0.31 & $0.6996 \pm 0.0036$ & $22.4231 \pm 0.1555$ & $0.7041$ & $22.6618$ \\
0.36 & $0.6967 \pm 0.0053$ & $22.3855 \pm 0.1151$ & $0.7049$ & $22.6301$ \\
0.41 & $0.6984 \pm 0.0033$ & $22.3294 \pm 0.1508$ & $0.7021$ & $22.6564$ \\
0.46 & $0.6947 \pm 0.0059$ & $22.3081 \pm 0.1781$ & $0.7049$ & $22.5967$ \\
0.51 & $0.6984 \pm 0.0045$ & $22.4265 \pm 0.1869$ & $0.7048$ & $\mathbf{22.7088}$ \\
0.56 & $0.6950 \pm 0.0056$ & $22.3109 \pm 0.1826$ & $0.7033$ & $22.6807$ \\
0.61 & $0.6908 \pm 0.0067$ & $22.1265 \pm 0.1720$ & $0.7014$ & $22.3127$ \\
0.66 & $0.6945 \pm 0.0082$ & $22.3099 \pm 0.2267$ & $0.7056$ & $22.6425$ \\
0.71 & $0.6932 \pm 0.0053$ & $22.2457 \pm 0.1427$ & $0.6989$ & $22.4198$ \\
\midrule

\multicolumn{5}{@{}l}{\textbf{Image: mrf\_n256\_l33\_i001}}\\
0.06 & $0.8994 \pm 0.0013$ & $30.5427 \pm 0.1616$ & $0.9017$ & $30.7903$ \\
0.11 & $0.8971 \pm 0.0034$ & $30.3894 \pm 0.1365$ & $0.9010$ & $30.5974$ \\
0.16 & $0.8964 \pm 0.0036$ & $30.4490 \pm 0.2195$ & $0.9013$ & $30.7278$ \\
0.21 & $0.8985 \pm 0.0055$ & $30.3760 \pm 0.3291$ & $0.9062$ & $30.8357$ \\
0.26 & $0.8987 \pm 0.0055$ & $30.4849 \pm 0.3438$ & $0.9059$ & $30.8721$ \\
0.31 & $\mathbf{0.9012 \pm 0.0037}$ & $\mathbf{30.5490 \pm 0.2889}$ & $0.9045$ & $30.9530$ \\
0.36 & $0.9001 \pm 0.0047$ & $30.5217 \pm 0.4981$ & $\mathbf{0.9077}$ & $\mathbf{31.3190}$ \\
0.41 & $0.9007 \pm 0.0048$ & $30.4412 \pm 0.3997$ & $0.9065$ & $31.0522$ \\
0.46 & $0.8979 \pm 0.0044$ & $30.4877 \pm 0.2599$ & $0.9070$ & $31.0410$ \\
0.51 & $0.8953 \pm 0.0046$ & $30.2034 \pm 0.2362$ & $0.9034$ & $30.5833$ \\
0.56 & $0.8994 \pm 0.0040$ & $30.3539 \pm 0.2287$ & $0.9057$ & $30.8360$ \\
0.61 & $0.8989 \pm 0.0058$ & $30.3975 \pm 0.3894$ & $0.9056$ & $30.8802$ \\
0.66 & $0.8969 \pm 0.0047$ & $30.3229 \pm 0.3326$ & $0.9038$ & $30.7874$ \\
0.71 & $0.8985 \pm 0.0041$ & $30.1684 \pm 0.3027$ & $0.9049$ & $30.6030$ \\
\midrule

\multicolumn{5}{@{}l}{\textbf{Image: mrf\_n512\_l9\_i001}}\\
0.06 & $0.8612 \pm 0.0026$ & $26.7903 \pm 0.1510$ & $0.8641$ & $26.9957$ \\
0.11 & $0.8610 \pm 0.0019$ & $26.7239 \pm 0.1166$ & $0.8630$ & $26.9087$ \\
0.16 & $0.8616 \pm 0.0034$ & $26.7570 \pm 0.2160$ & $0.8669$ & $26.9968$ \\
0.21 & $0.8598 \pm 0.0036$ & $26.6818 \pm 0.1649$ & $0.8655$ & $27.0004$ \\
0.26 & $\mathbf{0.8636 \pm 0.0027}$ & $\mathbf{26.8385 \pm 0.1727}$ & $\mathbf{0.8677}$ & $\mathbf{27.0684}$ \\
0.31 & $0.8619 \pm 0.0036$ & $26.7271 \pm 0.1872$ & $0.8666$ & $27.0331$ \\
0.36 & $0.8590 \pm 0.0032$ & $26.5669 \pm 0.2004$ & $0.8628$ & $26.8073$ \\
0.41 & $0.8548 \pm 0.0051$ & $26.3788 \pm 0.2136$ & $0.8607$ & $26.7051$ \\
0.46 & $0.8565 \pm 0.0035$ & $26.4718 \pm 0.2114$ & $0.8626$ & $26.8394$ \\
0.51 & $0.8562 \pm 0.0033$ & $26.4302 \pm 0.2015$ & $0.8615$ & $26.6638$ \\
0.56 & $0.8542 \pm 0.0036$ & $26.3494 \pm 0.2619$ & $0.8585$ & $26.7172$ \\
0.61 & $0.8494 \pm 0.0071$ & $26.1895 \pm 0.3391$ & $0.8620$ & $26.7437$ \\
0.66 & $0.8488 \pm 0.0045$ & $26.0763 \pm 0.2362$ & $0.8587$ & $26.5321$ \\
0.71 & $0.8464 \pm 0.0070$ & $25.9441 \pm 0.2300$ & $0.8591$ & $26.4245$ \\
\midrule

\multicolumn{5}{@{}l}{\textbf{Image: Blob}}\\
0.06 & $0.9884 \pm 0.0010$ & $30.0579 \pm 0.1702$ & $0.9898$ & $30.3163$ \\
0.11 & $0.9895 \pm 0.0011$ & $30.1937 \pm 0.1946$ & $0.9909$ & $30.4914$ \\
0.16 & $0.9900 \pm 0.0006$ & $30.3361 \pm 0.2568$ & $0.9910$ & $30.6989$ \\
0.21 & $0.9902 \pm 0.0007$ & $30.4132 \pm 0.1634$ & $0.9911$ & $30.6421$ \\
0.26 & $0.9909 \pm 0.0006$ & $30.6413 \pm 0.1610$ & $0.9917$ & $30.9527$ \\
0.31 & $0.9910 \pm 0.0007$ & $30.7072 \pm 0.2117$ & $0.9922$ & $31.0527$ \\
0.36 & $0.9916 \pm 0.0004$ & $30.7426 \pm 0.1284$ & $0.9922$ & $30.9559$ \\
0.41 & $0.9915 \pm 0.0003$ & $30.7220 \pm 0.1602$ & $0.9921$ & $31.0245$ \\
0.46 & $0.9920 \pm 0.0006$ & $30.9172 \pm 0.2821$ & $0.9930$ & $31.3742$ \\
0.51 & $0.9920 \pm 0.0004$ & $30.9161 \pm 0.2183$ & $0.9925$ & $31.1528$ \\
0.56 & $0.9921 \pm 0.0008$ & $30.9472 \pm 0.3417$ & $0.9934$ & $31.5669$ \\
0.61 & $0.9925 \pm 0.0006$ & $30.9968 \pm 0.3246$ & $0.9931$ & $\mathbf{31.6199}$ \\
0.66 & $0.9927 \pm 0.0006$ & $\mathbf{31.1771 \pm 0.2282}$ & $0.9935$ & $31.5255$ \\
0.71 & $\mathbf{0.9931 \pm 0.0006}$ & $31.1713 \pm 0.3133$ & $\mathbf{0.9938}$ & $31.5267$ \\
\midrule

\multicolumn{5}{@{}l}{\textbf{Image: Lincoln}}\\
0.06 & $0.9700 \pm 0.0022$ & $25.0408 \pm 0.1918$ & $0.9753$ & $25.5322$ \\
0.11 & $0.9731 \pm 0.0028$ & $25.4038 \pm 0.2962$ & $0.9778$ & $25.8434$ \\
0.16 & $0.9746 \pm 0.0022$ & $25.6306 \pm 0.2452$ & $0.9780$ & $26.0886$ \\
0.21 & $0.9776 \pm 0.0022$ & $25.9359 \pm 0.4291$ & $0.9820$ & $26.7767$ \\
0.26 & $0.9778 \pm 0.0016$ & $25.9491 \pm 0.2897$ & $0.9798$ & $26.3802$ \\
0.31 & $0.9777 \pm 0.0021$ & $25.9764 \pm 0.2787$ & $0.9811$ & $26.4948$ \\
0.36 & $0.9790 \pm 0.0019$ & $26.1633 \pm 0.2997$ & $0.9822$ & $26.6125$ \\
0.41 & $0.9788 \pm 0.0019$ & $26.1079 \pm 0.3469$ & $0.9816$ & $26.4724$ \\
0.46 & $\mathbf{0.9809 \pm 0.0015}$ & $26.4721 \pm 0.2952$ & $0.9830$ & $26.8248$ \\
0.51 & $0.9799 \pm 0.0021$ & $26.3493 \pm 0.2605$ & $0.9821$ & $26.7421$ \\
0.56 & $0.9799 \pm 0.0016$ & $26.2305 \pm 0.3264$ & $0.9833$ & $26.9814$ \\
0.61 & $0.9809 \pm 0.0024$ & $\mathbf{26.4877 \pm 0.3577}$ & $\mathbf{0.9850}$ & $\mathbf{27.2179}$ \\
0.66 & $0.9809 \pm 0.0014$ & $26.4376 \pm 0.2692$ & $0.9827$ & $26.6630$ \\
0.71 & $0.9806 \pm 0.0015$ & $26.4226 \pm 0.1980$ & $0.9825$ & $26.6461$ \\
\end{longtable}
}

\subsection{Experimental details}\label{sec:experimental_details}
Both algorithms are implemented in Julia (\cite{bezanson2017julia}) and were ran them on an HPE ProLiant DL380 running Debian GNU/Linux and equipped with two Intel Xeon 18-Core 5220 @2.2Ghz, two NVIDIA TESLA V100 32GB PCIe graphic cards, two SSD 480 GB hard drives, and 384GB RAM (12X32GB DDR4).
The single-spin-flip algorithm ran entirely on the CPU, whereas the PCA fully exploited the GPU. The GPU kernels have been implemented using CUDA.jl library (\cite{besard2018juliagpu}).

\section{Discussion of results and future developments}\label{sec:discussion}

The paper aims to assess whether PCA are a viable option to tackle inverse problems. 
In this respect, we believe that the obtained results support this hypothesis. 
In particular, we show that it may be worth losing the exact control of the stationary measure of the Markov chain in order to use an inherently parallel algorithm.
This work paves the road for PCA Markov Chain to the same type of evolution \emph{sequential} Markov Chain Monte Carlo followed over the last decades.
In this direction, we plan to further develop this research project along several lines

One of these lines concerns taking into account more realistic images together with more complex degradation types (such as multiplicative noise and combinations of noise and blur). To address these cases, different prior distributions should be considered. In particular, this task will require taking into account neighborhoods for the pixels that are more complex than the neighborhood of \eqref{eq:neighborhood} (visually represented in \Cref{fig:neighborhood}) and interactions that involve not only the values of neighboring pixels, but also the ``edges'' these contours create. 
Further, we should also consider an interaction between pixel that is more complex than that of
described by \eqref{eq:prior_pair_potential} (which is, essentially an \( L_{0} \) ``norm'') such as, for instance, \( L_{1} \) and \( L_{2} \) norms. Likewise, different norms should be considered for the inertial term of the posterior Hamiltonian \eqref{eq:denoising_pair_posterior_Hamiltonian}
of the PCA.

A second line of research should involve investigating the theoretical properties of the PCA, especially with respect to its mixing time and the conjectured concavity of the ``goodness of the reconstruction'' with respect to \( q \). In particular, providing rigorous (and useful) estimates of the chain's mixing time is a highly non-trivial task. Indeed, for large values of the inverse temperature \( \beta \) (supercritical regime), that is, when the stationary measure of the chain is concentrated on the global minima of the posterior Hamiltonian, the time that the chain requires to reach a global minimizer of the posterior Hamiltonian is essentially determined by the time it takes to leave the basin of attraction of the metastable states. To determine this time, a detailed analysis of the energy landscape induced by the posterior Hamiltonian would be required. Furthermore, in the supercritical regime, assessing the theoretical advantage of the PCA with respect to the sequential dynamics is a delicate matter (see \cite{louis2015supercritical}). Similarly, assessing theoretically the convergence of the equilibrium measure of the PCA and the target posterior distribution as \( q \to \infty\) is expected to be depend heavily on the precise characterization of the minimizers of the posterior Hamiltonian (see, e.g., \cite{procacci2016probabilistic, apollonio2019criticality}).
Nevertheless, from a practical point of view, the approach of a lazy PCA with a laziness parameter set heuristically has given encouraging results for instance in the context of combinatorial optimization (see \cite{apollonio2022shaken,fukushima2023mixing, isopi2024some, oknogi2022fully, scoppola2018gaussian,troiani2024probabilistic}).

Finally, we should put forward a comparison of the PCA approach with families of MCMC algorithms
such as Hamiltonian Monte Carlo and Langevin Monte Carlo. 


\section*{Acknowledgment}
{\small All the authors are members of the Gruppo Nazionale per l'Analisi Matematica, la Probabilità e le loro Applicazioni (GNAMPA) of the Istituto Nazionale di Alta Matematica (INdAM). D. Costarelli and M. Piconi are also members of the UMI (Unione Matematica Italiana) group T.A.A. (Teoria dell' Approssimazione e Applicazioni)}, and of the network RITA (Research ITalian network on Approximation).
The authors are thankful to the Department of Mathematics of the University of Rome ``Tor Vergata'' for providing access to computer resources.

\section*{Funding}
{\small The authors have been supported within the project PRIN 2022 PNRR: ``RETINA: REmote sensing daTa INversion with multivariate functional modeling for essential climAte variables characterization'', funded by the European Union under the Italian National Recovery and Resilience Plan (NRRP) of NextGenerationEU, under the Italian Minister of University and Research MUR (Project Code: P20229SH29, CUP: J53D23015950001). }

\section*{Author's contribution} 
\small{All authors contributed equally to this work for writing, reviewing, and editing. All authors have read and agreed to the published version of the manuscript.}

\section*{Conflict of interest/Competing interests}
\small{The authors declare that they have no conflict of interest and competing interests.}

\section*{Copyright}
{\small The MRF images were generated by the authors. The non-MRF images are available at the website \url{https://www.imageprocessingplace.com/DIP-3E/dip3e_book_images_downloads.htm} and do not require asking the permission of the authors if used for research purposes, as stated at the address \url{https://www.imageprocessingplace.com/downloads_V3/root_downloads/copyrights/dip3e_copyrights.htm}}.

\section*{Availability of data and material and Code availability}
{\small All the data generated for this study were stored in our laboratory and are not publicly available. Researchers who wish to access the data directly (including the ``.tif'' version of the images) can contact the corresponding author.}


\end{document}